\documentstyle[12pt]{article}
\def\b{\begin{equation}}
\def\e{\end{equation}}

\begin{document}
\thispagestyle{empty}
\begin{flushright}
FERMILAB-PUB-96/212-T\\
hep-ph/9608232\\
August 1996
\end{flushright}
\vspace*{2cm}
\centerline{\Large\bf Renormalization of}
\centerline{\Large\bf $\Delta B=2$ Transitions in the Static Limit}
\centerline{\Large\bf Beyond Leading Logarithms}
\vspace*{1.5cm}
\centerline{{\sc Gerhard Buchalla}} 
\bigskip
\centerline{\sl Theoretical Physics Department}
\centerline{\sl Fermi National Accelerator Laboratory}
\centerline{\sl P.O. Box 500, Batavia, IL 60510, U.S.A.}

\vspace*{3cm}
\centerline{\bf Abstract}
\vspace*{0.2cm}
\noindent 
The renormalization group evolution of $\Delta B=2$ transition
operators is studied at leading order in heavy quark effective
theory and at next-to-leading order in QCD. We calculate new
contributions that were not taken into account in previous work and
obtain the complete result to next-to-leading logarithmic accuracy.
The various scheme and scale dependences and their cancellation
in physical quantities are discussed explicitly.


\vfill
\newpage
 
\pagenumbering{arabic}
 
The phenomenon of $B_d-\bar B_d$ mixing is one of the most interesting
and important effects in flavordynamics. Most notably it is sensitive
to the mass and the weak coupling $V_{td}$ of the top quark, which play
an important role in testing the standard model picture of quark
mixing in general and of the unitarity triangle and its implications for
CP violation in particular. Unlike many other loop-induced
FCNC processes, $B_d-\bar B_d$ mixing has been observed and its
measurement provides some constraint on standard model parameters
already today. As our knowledge of the top quark mass increases and the
experimental result for the mixing parameter $x_d=\Delta m_{B_d}/\Gamma_B$
gains in accuracy, $B-\bar B$ mixing has the potential of providing very
valuable information on $|V_{td}|$. Such information is crucial for a
decisive test of our current understanding of flavordynamics.
\\
Unfortunately the analysis and interpretation of $B-\bar B$ mixing is
limited on the theoretical side through the need to evaluate the hadronic
matrix element of a four-quark operator between the $B$ meson states.
This computation involves QCD long distance dynamics, which pose a 
notoriously difficult problem not yet solved satisfactorily to date.
Ultimately the required matrix element should be calculable in the 
framework of lattice gauge theory.
However, since the b-quark mass is larger than currently attainable
lattice cutoffs, it is difficult to treat $B$ mesons with fully
dynamical quarks in numerical simulations. In this situation it is
useful to employ heavy quark effective theory (HQET). This approach
allows one to extract the dependence of the matrix element on the heavy
mass $m_b$ analytically, leaving only light degrees of freedom as
dynamical variables in a lattice calculation.
\\
It is the purpose of this Letter to discuss the relation of the full
theory operator to its HQET counterparts to leading order in the
$1/m$ expansion and to next-to-leading order in renormalization group
improved perturbation theory, thereby generalizing, repectively
completing results that have already been obtained in the literature
\cite{FHH,GIM}

The effective hamiltonian describing $B_d-\bar B_d$ mixing can be
written as
\begin{equation}\label{heff}
{\cal H}^{\Delta B=2}_{eff}=\frac{G^2_F}{16\pi^2}M^2_W
(V^\ast_{tb}V_{td})^2 C(x_t(\mu_t),\mu_t,\mu_b) Q + h.c.
\end{equation}
with
\begin{equation}\label{qdef}
Q=(\bar bd)_{V-A}(\bar bd)_{V-A}
\end{equation}
and $x_t=m^2_t/M^2_W$. From this hamiltonian the $B_d-\bar B_d$ mixing
parameter $x_d=\Delta m_{B_d}/\Gamma_B$ can be derived with the result
\begin{equation}\label{xdef}
x_d=\tau_B \frac{G^2_F}{16\pi^2}\frac{M^2_W}{m_B}|V_{td}|^2
C(x_t(\mu_t),\mu_t,\mu_b) \langle Q(\mu_b)\rangle
\end{equation}
where
\begin{equation}\label{qmeb}
\langle Q(\mu_b)\rangle\equiv\langle\bar B|Q(\mu_b)|B\rangle=
\frac{8}{3}f^2_B m^2_B B_B(\mu_b)
\end{equation}
is the required hadronic matrix element mentioned above. $f_B$ is the
B-meson decay constant in the normalization in which $f_\pi=131 MeV$
and $B_B$ is the bag parameter, defined through (\ref{qmeb}).
\\
The effective hamiltonian formalism is based on the operator product
expansion, which achieves a factorization of short-distance and
long-distance contributions to the amplitude in (\ref{xdef}).
The dependence on the top-quark mass and the short-distance QCD effects
from the high matching scale $\mu_t={\cal O}(m_t)={\cal O}(M_W)$, at
which top and $W$ are integrated out, down to a scale $\mu_b={\cal O}(m_b)$
are contained in the Wilson coefficient function $C(x_t(\mu_t),\mu_t,\mu_b)$.
This coefficient is calculable in RG improved perturbation theory. It is
independent of the scale $\mu_t$ to the considered order in $\alpha_s$.
On the other hand $C$ does depend on the low scale $\mu_b$ and on the
renormalization scheme chosen to subtract divergences in its calculation.
Both dependences are canceled by the hadronic matrix element in the
product $C\cdot\langle Q\rangle$. The important feature of the scale
and scheme dependence and their cancellation appears only at
${\cal O}(\alpha_s)$ and hence can be meaningfully addressed only beyond
the leading logarithmic approximation. For the coefficient function
$C(x_t(\mu_t),\mu_t,\mu_b)$ the next-to-leading order analysis has been 
performed and discussed in detail in \cite{BJW}. This calculation resums
the leading $(\alpha_s^n\ln^n\mu_t/\mu_b)$ and next-to-leading
$(\alpha_s^{n+1}\ln^n\mu_t/\mu_b)$ logarithmic QCD corrections to all orders
$n=0$, $1$, $2$, $\ldots$ in perturbation theory and eliminates the
$\mu_t$-dependence to order ${\cal O}(\alpha_s)$, which considerably
reduces the theoretical uncertainty in the short-distance calculation.
In fact the $\mu_t$-dependence is entirely negligible after the inclusion
of NLO corrections \cite{BJW}.

Whereas the coefficient $C$ incorporates the short-distance QCD effects
from scale $\mu_t$ down to $\mu_b$, the matrix element $\langle Q\rangle$
contains the "long-distance" contributions below $\mu_b={\cal O}(m_b)$.
However, since QCD continues to be perturbative from scales $\mu_b$
down to $\mu\sim 1 GeV$, the corresponding contributions to the matrix
element may still be extracted in an analytical way using the framework
of HQET. Within this formalism the full theory matrix element
$\langle Q(\mu_b)\rangle$ may be expanded as follows \cite{FHH}
\begin{equation}\label{qc12t}
\langle Q(\mu_b)\rangle=C_1(\mu_b, \mu)\langle\tilde Q(\mu)\rangle+
 C_2(\mu_b, \mu)\langle\tilde Q_s(\mu)\rangle +{\cal O}(\frac{1}{m})
\end{equation}
where we are working to leading order in $1/m$. In this order two
operators appear in the effective theory. They are given by
\begin{equation}\label{qqst}
\tilde Q=2(\bar h^{(+)}d)_{V-A}(\bar h^{(-)}d)_{V-A}\qquad\qquad
\tilde Q_s=2(\bar h^{(+)}d)_{S-P}(\bar h^{(-)}d)_{S-P}
\end{equation}
with $(\bar hd)_{V-A}\equiv\bar h\gamma^\mu(1-\gamma_5)d$ and
$(\bar hd)_{S-P}\equiv\bar h(1-\gamma_5)d$. Here the heavy quark 
fields are defined in the usual way as
\begin{equation}\label{hvb}
h^{(\pm)}(x)=e^{\pm imv\cdot x}\frac{1\pm\not\! v}{2} b(x)
\end{equation}
$\tilde Q$, $\tilde Q_s$ are the HQET operators relevant for the case of a
$B\to\bar B$ transition. The field $\bar h^{(+)}$ creates a heavy quark,
while $\bar h^{(-)}$ annihilates a heavy antiquark. Since the
effective theory field $\bar h^{(+)}$ ($\bar h^{(-)}$) cannot, unlike the
full theory field $\bar b$ in $Q$ (\ref{qdef}), at the same time annihilate (create)
the heavy antiquark (heavy quark), explicit factors of two have to appear
in (\ref{qqst}). Due to the mass of the $b$ quark an additional
operator $\tilde Q_s$ with scalar--pseudoscalar structure is generated
in the effective theory.
\\
(\ref{qc12t}) has the form of an operator product expansion. The 
short-distance contributions of scales between $\mu_b$ and $\mu$ are
factorized into the coefficients $C_{1,2}$, while 
$\langle\tilde Q(\mu)\rangle$ and $\langle\tilde Q_s(\mu)\rangle$ contain
the contributions below $\mu$. In particular the leading dependence of
$\langle Q(\mu)\rangle$ on the $b$-quark mass is fully separated into the
coefficient functions and no reference to $m_b$ is present in the
effective theory matrix elements, which makes them more readily
computable by lattice methods than $\langle Q(\mu)\rangle$.
\\
In a first step the coefficients $C_{1,2}$ have to be determined at
$\mu=\mu_b$ by calculating the matching of the full theory operator $Q$
onto its HQET counterparts $\tilde Q$, $\tilde Q_s$ to ${\cal O}(\alpha_s)$
in perturbation theory. After performing a proper factorization of long-
and short-distance contributions, the coefficient functions are independent
of the infrared properties of the amplitude (\ref{qc12t}), in particular
of the treatment of external lines. Therefore, for the determination
of the coefficients, the arbitrary external states may be chosen
according to calculational convenience. The simplest technique is to put
the heavy quark lines on shell, to set all other external momenta to zero
and to regularize both ultraviolet and infrared divergences dimensionally.
In this case the ${\cal O}(\alpha_s)$ corrections to the matrix elements
of $\tilde Q$, $\tilde Q_s$ vanish. Using this method, performing the 
necessary renormalizations in the $\overline{MS}$ scheme with
anticommuting $\gamma_5$ (NDR scheme) and subtracting evanescent
contributions in the usual way \cite{BW}, we obtain
\begin{equation}\label{c1mub}
C_1(\mu_b,\mu_b)=1+\frac{\alpha_s(\mu_b)}{4\pi}
\left[(\tilde\gamma^{(0)}_{11}-\gamma^{(0)})\ln\frac{\mu_b}{m_b}+
\tilde B-B\right]
\end{equation}
\begin{equation}\label{c2mub}
C_2(\mu_b,\mu_b)=\frac{\alpha_s(\mu_b)}{4\pi} \tilde B_s
\end{equation}
where
\begin{equation}\label{gtcf}
\gamma^{(0)}=6\frac{N-1}{N}\qquad\qquad
\tilde\gamma^{(0)}_{11}=-6C_F\qquad\qquad
C_F=\frac{N^2-1}{2N}
\end{equation}
\begin{equation}\label{bbst}
\tilde B-B=-\frac{8N^2+9N-15}{2N}\qquad B=11\frac{N-1}{2N}\qquad
\tilde B_s=-2(N+1)
\end{equation}
and N is the number of colors. For $N=3$ this result coincides
with the calculation of \cite{FHH}.The logarithmic term in (\ref{c1mub})
reflects the ${\cal O}(\alpha_s)$ scale dependence of the matrix
elements of $Q$ and $\tilde Q$. Accordingly its coefficient is given by
the difference in the one-loop anomalous dimensions of these
operators, $\gamma^{(0)}$ and $\tilde\gamma^{(0)}_{11}$. At the same time
this logarithm expresses the complete dependence of the full theory
matrix element $\langle Q\rangle$ on the $b$-quark mass, to 
${\cal O}(\alpha_s)$ and to leading order in $1/m$, which is thus
factorized into the short-distance coefficient function $C_1$. Besides
the logarithm (\ref{c1mub}) contains a scheme dependent constant. We have
written this term in the form $\tilde B-B$ in order to make the
cancellation of scheme dependences, to be discussed below, more transparent.
Here $B$ is a constant characterizing the scheme dependence of the full theory
matrix element $\langle Q\rangle$, given here in the NDR scheme
(see also \cite{BJW}).

In a next step we want to evolve the coefficients $C_{1,2}$ from the 
$b$-scale $\mu_b={\cal O}(m_b)$ down to a typical hadronic scale
$\mu\sim 1GeV$ in next-to-leading order of RG improved perturbation
theory. In this way we are resumming to all orders 
the leading $(\alpha_s^n\ln^n\mu_b/\mu)$ and next-to-leading
$(\alpha_s^{n+1}\ln^n\mu_b/\mu)$ logarithmic contributions, which are of
${\cal O}(1)$ and ${\cal O}(\alpha_s)$, respectively, in the limit
$\mu_b\gg \mu$, where the logarithms become large.
\\
The evolution of the coefficients $\vec C^T=(C_1, C_2)$ is described
by the RG equation
\begin{equation}\label{rgct}
\frac{d}{d\ln\mu}\vec C(\mu_b,\mu)=\tilde\gamma^T \vec C(\mu_b,\mu)
\end{equation}
where the matrix of anomalous dimensions of the operators $\tilde Q$,
$\tilde Q_s$ can be written to two-loop accuracy as
\begin{equation}\label{gat}
\tilde\gamma=\frac{\alpha_s}{4\pi}
\left(\begin{array}{cc} \tilde\gamma^{(0)}_{11} & 0 \\
                        \tilde\gamma^{(0)}_{21} & \tilde\gamma^{(0)}_{22}
\end{array}\right)
+\left(\frac{\alpha_s}{4\pi}\right)^2
\left(\begin{array}{cc} \tilde\gamma^{(1)}_{11} & 0 \\
                        \tilde\gamma^{(1)}_{21} & \tilde\gamma^{(1)}_{22}
\end{array}\right)
\end{equation}
The first row of $\tilde\gamma$ has already been considered to
next-to-leading order in \cite{GIM} and we will use the results obtained in
this paper here. The vanishing of the (12)-entry in $\tilde\gamma$
means that $\tilde Q$ does not mix into the operator $\tilde Q_s$.
The second row of $\tilde\gamma$ will be the subject of the present
article.
\\
Using the well known general expressions for the solution of (\ref{rgct})
at next-to-leading order \cite{BJLW}, the structure of $\tilde\gamma$
given in (\ref{gat}) and the initial values of $C_{1,2}$ from
(\ref{c1mub}), (\ref{c2mub}) we obtain after inserting the solution
$\vec C(\mu_b,\mu)$ into (\ref{qc12t})
\begin{eqnarray}\label{qqtg}
\lefteqn{\langle Q(\mu_b)\rangle=} \nonumber \\
& & \left[\frac{\alpha_s(\mu_b)}{\alpha_s(\mu)}\right]^{\tilde d_1}
\left(1+\frac{\alpha_s(\mu_b)}{4\pi}
\left[(\tilde\gamma^{(0)}_{11}-\gamma^{(0)})\ln\frac{\mu_b}{m_b}+\tilde B-B
-\tilde J\right]+\frac{\alpha_s(\mu)}{4\pi}\tilde J\right)
\langle\tilde Q(\mu)\rangle  \nonumber \\
& & +\frac{\alpha_s(\mu_b)}{4\pi}
\frac{\tilde\gamma^{(0)}_{21}}{\tilde\gamma^{(0)}_{11}-\tilde\gamma^{(0)}_{22}}
\tilde B_s\left(
\left[\frac{\alpha_s(\mu_b)}{\alpha_s(\mu)}\right]^{\tilde d_1}-
\left[\frac{\alpha_s(\mu_b)}{\alpha_s(\mu)}\right]^{\tilde d_2}\right)
\langle\tilde Q(\mu)\rangle  \nonumber \\
& & +\frac{\alpha_s(\mu_b)}{4\pi}
\left[\frac{\alpha_s(\mu_b)}{\alpha_s(\mu)}\right]^{\tilde d_2}
\tilde B_s \langle\tilde Q_s(\mu)\rangle
\end{eqnarray}
Here
\begin{equation}\label{d12j}
\tilde d_1=\frac{\tilde\gamma^{(0)}_{11}}{2\beta_0}\qquad
\tilde d_2=\frac{\tilde\gamma^{(0)}_{22}}{2\beta_0}\qquad
\tilde J=\frac{\tilde d_1}{\beta_0}\beta_1-
 \frac{\tilde\gamma^{(1)}_{11}}{2\beta_0}
\end{equation}
where
\begin{equation}\label{b0b1}
\beta_0=\frac{11N-2f}{3}\qquad\qquad
\beta_1=\frac{34}{3}N^2-\frac{10}{3}N f-2 C_F f
\end{equation}
are the beta-function coefficients with $N$ ($f$) the number of
colors (flavors).
\\
Eq. (\ref{qqtg}) represents the relation between the matrix element of
the full theory operator $Q$, normalized at the "high" scale $\mu_b$,
and the matrix elements of the leading HQET operators $\tilde Q$ and
$\tilde Q_s$, normalized at a low scale $\mu\sim 1 GeV$. It is valid to
leading order in the $1/m_b$ expansion and to next-to-leading order in
RG improved QCD perturbation theory. This leading contribution in the
heavy quark expansion corresponds to the static limit for the $b$-quark.
Going beyond this approximation requires the consideration of several
new operators, which arise at the next order in $1/m$. These
contributions have been studied in \cite{KM} in the leading
logarithmic approximation.
\\
We further remark that (\ref{qqtg}) as it stands is valid in the
continuum theory. In order to use lattice results one still has to
perform an ${\cal O}(\alpha_s)$ matching of $\tilde Q$ to its
lattice counterpart. This step however does not involve any further
renormalization group improvement since by means of (\ref{qqtg})
$\langle\tilde Q\rangle$ is already normalized at the appropriate low
scale $\mu$. For $\tilde Q_s$, whose coefficient is ${\cal O}(\alpha_s)$,
a tree level matching is sufficient.
\\
Note that due to the particular form of the RG evolution and the fact that
$C_2$ is only ${\cal O}(\alpha_s)$, the anomalous dimension coefficients
$\tilde\gamma^{(1)}_{21}$ and $\tilde\gamma^{(1)}_{22}$ in (\ref{gat}) do
not appear in (\ref{qqtg}) and hence are irrelevant at NLO.
The one-loop anomalous dimension $\tilde\gamma^{(0)}_{11}$, describing
the leading logarithmic evolution of the effective theory operator
matrix element $\langle\tilde Q\rangle$, has been first calculated in
\cite{VS} and \cite{PW} with the result quoted in (\ref{gtcf}). The
computation of the two-loop anomalous dimension was performed in
\cite{GIM}, who obtains in the NDR scheme
\begin{equation}\label{gt11}
\tilde\gamma^{(1)}_{11}=
 -\frac{N-1}{12N}\left[127 N^2+143 N+63-\frac{57}{N}+
\left(8 N^2-16N+\frac{32}{N}\right)\pi^2-(28N+44)f\right]   
\end{equation}
As (\ref{qqtg}) shows, the complete NLO expression requires in addition to
$\tilde\gamma^{(0)}_{11}$ and $\tilde\gamma^{(1)}_{11}$ the one-loop
anomalous dimensions
$\tilde\gamma^{(0)}_{22}$ and $\tilde\gamma^{(0)}_{21}$, which describe the
evolution of the operator $\tilde Q_s$ and the mixing of $\tilde Q_s$
into $\tilde Q$, respectively. We have calculated these quantities and find
\begin{equation}\label{g2122}
\tilde\gamma^{(0)}_{21}=\frac{N+1}{N}\qquad\qquad
\tilde\gamma^{(0)}_{22}=-\frac{3N^2-4N-7}{N}
\end{equation}
The determination of
$\tilde\gamma^{(0)}_{21}$ and $\tilde\gamma^{(0)}_{22}$ is
straightforward (for general reviews see \cite{NEU,BBL}). 
We only mention that in order to reexpress additional
operator structures, arising in the calculation, in terms of the basis
operators $\tilde Q$, $\tilde Q_s$, we made use of the identity
(valid in four dimensions for $v^2=1$)
\begin{equation}\label{vvid}
\not\! v\gamma^\mu\gamma^\nu(1-\gamma_5)\otimes\not\! v
\gamma_\mu\gamma_\nu(1-\gamma_5)\equiv
4\gamma^\nu(1-\gamma_5)\otimes\gamma_\nu(1-\gamma_5)
\end{equation}
and of the projection property $\not\! v h^{(\pm)}=\pm h^{(\pm)}$ of the
heavy quark effective fields (\ref{hvb}). The relation (\ref{vvid}) is
also necessary for the matching calculation leading to
(\ref{c1mub}), (\ref{c2mub}).
\\
Let us discuss the structure of relation (\ref{qqtg}) in more detail.
In particular we would like to consider the scale
and scheme dependences present in (\ref{qqtg}) and to clarify their
cancellation in physical quantities. The dependence on $\mu$
in the first factor on the r.h.s. of (\ref{qqtg}) is canceled by the
$\mu$-dependence of $\langle\tilde Q(\mu)\rangle$. The dependence on
$\mu_b$ of this factor is canceled by the explicit $\ln\mu_b$ term
proportional to $\tilde\gamma^{(0)}_{11}$. Hence the only scale dependence
remaining on the r.h.s., to the considered order ${\cal O}(\alpha_s)$, 
is the one $\sim\alpha_s(\mu_b)\gamma^{(0)}\ln\mu_b$. This is
precisely the scale dependence of the full theory matrix element on the
l.h.s., which is required to cancel the corresponding dependence of
the Wilson coefficient in (\ref{xdef}). Similarly the term $\sim\alpha_s(\mu_b)B$
represents the correct scheme dependence of $\langle Q(\mu_b)\rangle$,
while the scheme dependence of $\alpha_s(\mu)\tilde J$ cancels with the
one of $\langle\tilde Q(\mu)\rangle$ and the difference
$\tilde B-\tilde J$ is scheme independent by itself.
Likewise the remaining contributions involving $\tilde B_s$ are scale and
scheme independent to the considered order.
\\
The relation (\ref{qqtg}) involves two different renormalization scales,
the scale $\mu$ at which the HQET matrix elements are defined and the
scale $\mu_b$ where the transition from full QCD to HQET is made. As is
evident from our discussion, the precise values of these scales are
arbitrary as long as $\mu={\cal O}(1 GeV)$ and $\mu_b={\cal O}(m_b)$,
where the hierarchy $\mu\ll\mu_b$ is assumed.
However, since the logarithms $\ln\mu_b/\mu$ are not really very large
in the case at hand, one might consider neglecting higher order
resummations of logarithms altogether and content oneself with the strict
$\alpha_s$-corrections alone. This limit can be obtained by expanding
(\ref{qqtg}) to first order in $\alpha_s$. In this case the expression
(\ref{qqtg}) simplifies to the result used in \cite{FHH} when $\mu_b$
is set equal to $\mu$. This approximation is fully consistent. 
Yet it is useful and interesting to
have the more complete expression (\ref{qqtg}) at hand, which
allows one to quantify the effects of the leading and next-to-leading
logarithms $\ln\mu_b/\mu$ in this relation.

Numerically, for $\mu_b=4.8GeV$, $\mu=1GeV$,
$\Lambda^{(4)}_{\overline{MS}}=0.3 GeV$ and $f=4$ flavors,
the coefficient of 
$\alpha_s(\mu_b)/(4\pi) \tilde B_s\langle\tilde Q_s(\mu)\rangle$
in (\ref{qqtg}) reads 1.13, instead of 1 if the summation of
logarithms was neglected. The new contribution proportional
to $\tilde B_s$ to the coefficient of
$\langle\tilde Q(\mu)\rangle$ is about $+0.01$.
Although these effects are not very large, they have to be
taken into account for consistency in a complete analysis within the
next-to-leading logarithmic approximation.
 
\vspace*{0.4cm}

{\bf Note added:} After completion of this work we became aware of
reference \cite{CFG} in which the same topic is addressed.
The results of \cite{CFG} are in agreement with our findings.

\vfill\eject
 
\end{document}